# Track Before Detect of Low SNR Objects in a Sequence of Image Frames Using Particle Filter

Reza Rezaie[1]

**Abstract**- A multiple model track-before-detect (TBD) particle filter-based approach for detection and tracking of low signal to noise ratio (SNR) objects based on a sequence of image frames in the presence of noise and clutter is briefly studied in this letter. At each time instance after receiving a frame of image, first, some preprocessing approaches are applied to the image. Then, it is sent to the multiple model TBD particle filter for detection and tracking of an object. Performance of the approach is evaluated for detection and tracking of an object in different scenarios including noise and clutter.

## I. INTRODUCTION

There are several methods for detection and tracking of low observable targets. In conventional methods, first, target is detected based on thresholding the received image and then its position can be estimated with a higher accuracy using a tracking filter (internal boxes in figure (1)). However, if the SNR of the target is not high enough to be always detected based on one frame thresholding, several consecutive frames of measurements can be used to detect the target using signal accumulation over time. In other word, such methods try to postpone thresholding to accumulate more power from the target to be able to detect it with a higher probability of detection. Therefore, in such methods detection and tracking must be done simultaneously (large box in figure (1)).

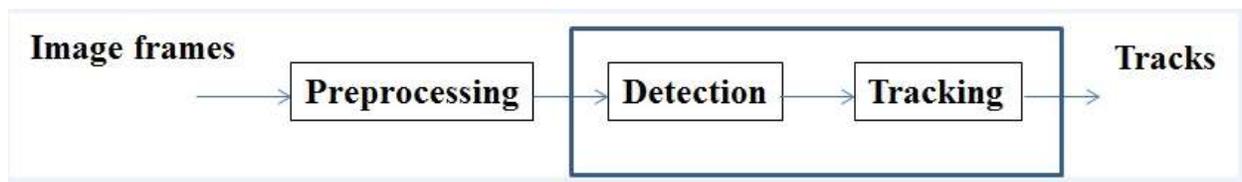

Fig (1): Different methods for detection and tracking of low observable targets

A TBD particle filter detection and tracking of low observable targets was presented in [3]. There is no maneuvering handling logic in the TBD approach of [3].

---

[1] rrezaie@uno.edu, rezarezaie01@gmail.com

First, some required preprocessing is briefly discussed. Then, the TBD detection and tracking approach of [3] is reviewed. A multiple model extension of the TBD approach of [3] is derived to handle low observable maneuvering targets and is used in simulations. Details of derivation and formulation of the derived multiple model TBD is skipped. Simulations show the performance of the multiple model TBD particle filter.

## II. PREPROCESSING

To match the image frames to the assumptions of the detection and tracking algorithm, some preprocessing is required as follows.

**Inverse Filtering**

The considered detection and tracking approach assumes that the target is a point target and not an extended one. So, if the target in the received frame of image is an extended one, it can be converted to a point target using inverse filtering. Conceptually, an extended target can be understood as a point target affected by a filter. So, to restore the target as a point one, we apply the inverse filtering (block diagram below). An example is shown in figure (2).

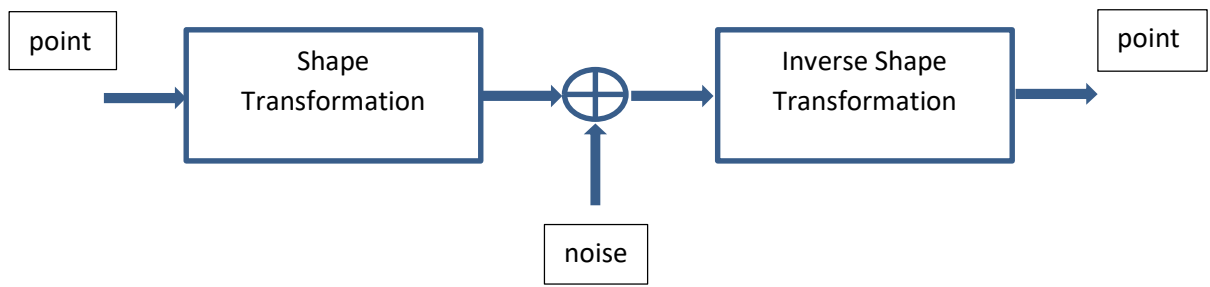

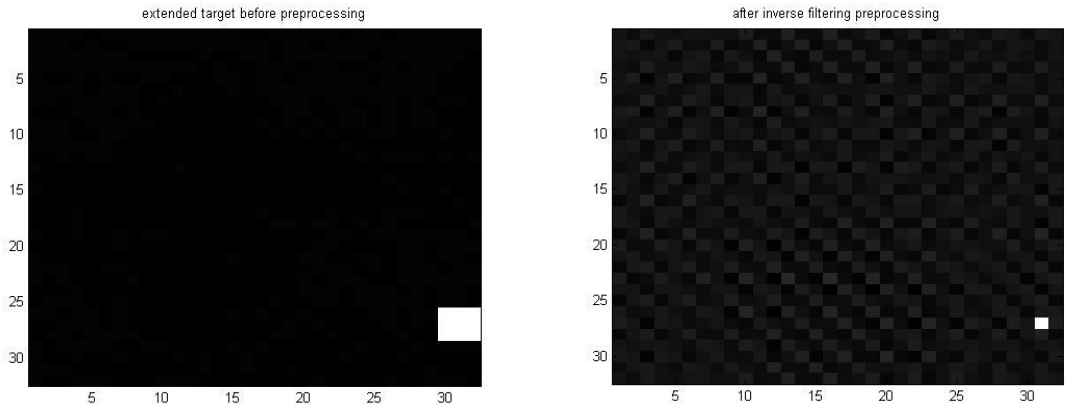

**Fig (2): (left) An extended target (3*3 dimension), (right) point target after applying inverse filtering and some preprocessing**

**Clutter Suppression**

If there is any clutter in the image frames the detection and tracking algorithm cannot perform well. So, in the preprocessing part it is required to suppress the clutter. The clutter suppression is done based on background subtraction since it is assumed that the movement of the clutter is less than targets and it is negligible.

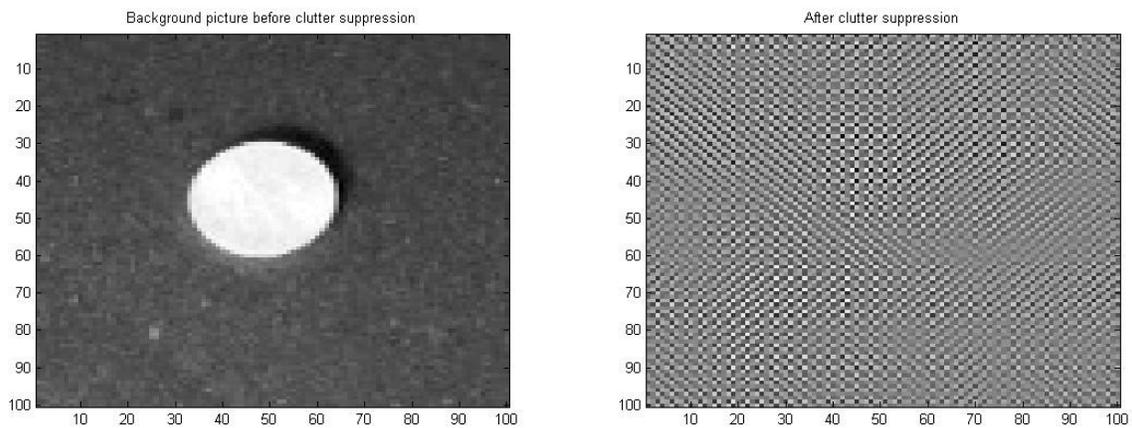

**Fig (3): (left) Background with clutter before clutter suppression, (right) background after clutter suppression and some preprocessing**

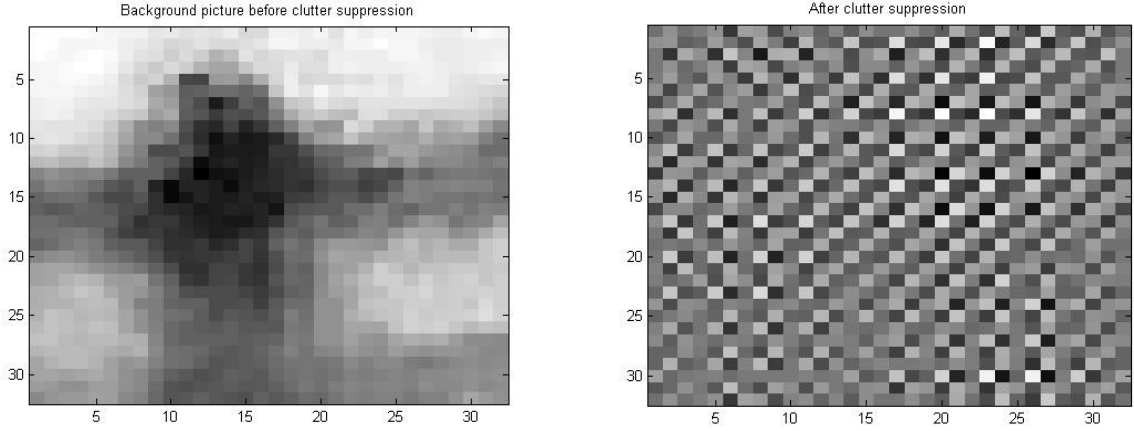

Fig (4): (left) Background with clutter before clutter suppression, (right) background after clutter suppression and some preprocessing

**Estimate of Variance**

The variance of noise must be known for the detection and tracking algorithm. Without clutter, a maximum likelihood (ML) estimate of the variance of the noise can be calculated based on a frame. However, in presence of clutter the ML estimate of the variance of noise is calculated based on the processed image after clutter suppression.

**Target Intensity Estimation in Presence of Intensity Fluctuations**

To deal with target intensity fluctuations, object intensity is augmented to the state vector as a state variable to be estimated with the help of particle filter along the algorithm.

## III. DETECTION AND TRACKING

**Target Model**

Target dynamic model is as follows

$$x_k = F x_{k-1} + v_k \qquad F = \begin{bmatrix} 1 & T & 0 & 0 \\ 0 & 1 & 0 & 0 \\ 0 & 0 & 1 & T \\ 0 & 0 & 0 & 1 \end{bmatrix} \qquad (1)$$

In which x is the state vector including position and velocity in 2D space, and v is dynamic noise, and F is the transition matrix.

Target may appear or disappear any time in the space. A variable is defined with Markov model for presence of target as follows with 1 indicating presence and 0 indicating absence of a target [1]-[10].

$$E_k \in \{0,1\} \qquad P_b = \Pr\{E_k = 1 | E_{k-1} = 0\} \qquad P_d = \Pr\{E_k = 0 | E_{k-1} = 1\}$$

Therefore,

$$\Pi_E = \begin{bmatrix} 1-P_b & P_b \\ P_d & 1-P_d \end{bmatrix}$$

And the initial probability at the beginning is

$$\mu_1 = \Pr\{E_1 = 1\}$$

It is assumed that these probabilities are known, otherwise they can be estimated.

**Sensor Model**

The received intensity in each pixel is modeled as follows

$$z_k^{(i,j)} = \begin{cases} h_k^{(i,j)}(x_k) + w_k^{(i,j)} & \text{if target present} \\ w_k^{(i,j)} & \text{if target absent} \end{cases} \qquad (2)$$

In which

$$h_k^{(i,j)}(x_k) = \begin{cases} I_k, & \text{if target is in cell (i, j)} \\ 0, & \text{otherwise} \end{cases}$$

$\Delta_x \times \Delta_y$ : Cell dimension $(i\Delta_x, j\Delta_y)$, $i = 1,...,n$, $j = 1,...,m$

$w_k^{(i,j)} : N(0, \sigma^2)$ (Observation noise in pixel (i, j) at time k)

$I_k$ : Target Intensity

And the observation of each image frame is as follows

$$Z_k = \{z_k^{(i,j)} : i = 1,...,n, \; j = 1,...,m\}$$

And all the image frames since the beginning to time k is denoted as

$$Z^k = \{Z_i, i = 1,...,k\}$$

**Detection and Tracking**

This method estimates the joint density of dynamic state and presence probability [1]-[10]. Then, if based on the presence probability estimate it is decided that there is a target in the space, target state vector can be estimated based on output of the filter. The presence probability also is calculated based on the output of the filter at every time instance. Joint density of presence probability and target state vector based on all the observations since the beginning to the current time presented in [3] is as follows

$$p(x_k, E_k | Z^k) = p(x_k | E_k, Z^k) p(E_k | Z^k) \quad (3)$$

Since density of state vector is calculated only if target is present, and since sum of presence and absence probabilities is 1, it is enough to recursively calculate the following terms

$$p(x_k | E_k = 1, Z^k)$$

$$p(E_k = 1 | Z^k)$$

Recursive calculation of the above two terms can be found in [3] and is provided in Appendix.

The above approach of [3] does not have any logic to handle maneuvering targets. We derived a multiple model extension of the above TBD approach to handle detection and tracking of low observable maneuvering targets. The derived multiple model TBD is based on

$$p(x_k, r_k, E_k | Z^k) = p(x_k, r_k | E_k, Z^k) p(E_k | Z^k)$$

and similar to the above, it is enough to recursively calculate the following two terms

$$p(x_k, r_k | E_k = 1, Z^k)$$

$$p(E_k = 1 | Z^k)$$

We skip the details of the derivation and the formulation. Later, we implement the multiple model TBD using particle filter and use it in simulations.

## IV. APPLYING TBD DETECTION AND TRACKING ALGORITHM USING PARTICLE FILTER

The multiple model TBD approach, pointed out in the previous section, is implemented using a particle filter and is used in simulations. We skip the details of its formulation.

The TBD method of [3] can be implemented using a particle filter. The TBD particle filter of [3] is presented below.

Since the algorithm is recursive, assume that $N_c$ particles $\{x_{k-1}^i | i=1,...,N\}$ are available from previous time describing the density function of state vector corresponding to the last time (Note that the weights are equal after resampling). Also, it is assumed that the estimate of the probability of target presence corresponding to the last time is available. Then, the recursive algorithm for detection and tracking of target is as follows

- Newborn particles are generated based on a proposal function (which can be a uniform density over those spots of the space with higher probability of including target)

$$x_k^{(b)i} \sim q(x_k | E_k = 1, E_{k-1} = 0, Z_k) \qquad (4)$$

The corresponding weights are calculated as follows

$$\widetilde{w}_k^{(b)i} = \frac{L(z_k | x_k^{(b)i}, E_k^{(b)i} = 1) p(x_k^{(b)i} | E_k^{(b)i} = 1, E_{k-1}^{(b)i} = 0)}{N_b q(x_k^{(b)i} | E_k^{(b)i} = 1, E_{k-1}^{(b)i} = 0, z_k)} \qquad (5)$$

And after normalizing

$$w_k^{(b)i} = \frac{\widetilde{w}_k^{(b)i}}{\sum_{j=1}^{N_b} \widetilde{w}_k^{(b)j}} \qquad (6)$$

- Continuing (survival) particles are generated. The considered proposal function is the dynamic model of target

$$\widetilde{w}_k^{(c)i} = \frac{1}{N_c} L(z_k | x_k^{(c)i}, E_k^{(c)i} = 1) \qquad (7)$$

$$w_k^{(c)i} = \frac{\tilde{w}_k^{(c)i}}{\sum_{j=1}^{N_c} \tilde{w}_k^{(c)j}} \qquad (8)$$

- Following coefficients are calculated for the calculation of target presence probability

$$\tilde{M}_b = p_b [1 - p_{k-1}] \sum_{i=1}^{N_b} \tilde{w}_k^{(b)i} \qquad (9)$$

$$\tilde{M}_c = [1 - p_d] p_{k-1} \sum_{i=1}^{N_b} \tilde{w}_k^{(c)i} \qquad (10)$$

After normalization

$$M_b = \tilde{M}_b / (\tilde{M}_b + \tilde{M}_c) \qquad (11)$$

$$M_c = \tilde{M}_c / (\tilde{M}_b + \tilde{M}_c) \qquad (12)$$

- Probability of presence at current time

$$p_k = \frac{\tilde{M}_b + \tilde{M}_c}{\tilde{M}_b + \tilde{M}_c + p_d p_{k-1} + [1 - p_b][1 - p_{k-1}]} \qquad (13)$$

Also, these weights are computed for calculation of posterior density of target state vector.

$$\hat{w}_k^{(b)i} = M_b w_k^{(b)i} \qquad (14)$$

$$\hat{w}_k^{(c)i} = M_c w_k^{(c)i} \qquad (15)$$

Then two sets of particles (newborn and continuation) are considered together

$$\{(x_k^{(t)i}, \hat{w}_k^{(t)i}) | i = 1,..., N_t, t = c, b\} \qquad (16)$$

- Finally, resampling is applied to the whole particles so that $N_c + N_b$ particles are reduced to $N_c$ particles.

After doing these steps, $\{(x_k^i, \frac{1}{N_c}) | i=1,...,N_c\}$ particles estimate density function of the state vector and the probability of target presence. Then, comparing this probability with a threshold (e.g., 0.6), the algorithm decides if there is a target in the space or not.

In the next section, this algorithm is applied to different scenarios to illustrate its performance in different situations.

## V. SIMULATIONS

To illustrate the performance of the multiple model TBD approach in detection and tracking, different scenarios have been considered including noise, clutter, synthetic image, real image, point and extended target with/without rotation.

**Scenario 1**: **Detection and tracking of a maneuvering point target in noise and clutter background (synthetic image)**

In this scenario the mean of the intensity of target is about 5, and the variance of noise is 1 (6.5 dB SNR before preprocessing). The intensity of clutter is 10 which is much more than that of target. Markov model transition probabilities are 0.05 and 0.95.

As it can be seen in the figure (4), (5), and (6) when there is no target in the space particles are uniformly spread in the space. We consider a very low threshold at the very beginning because pixels with very low intensity are not desired. This is the reason that particles are not completely uniformly spread in the space). When there is a target in the space, according to noise and target intensity realizations, the target may/may not be visible, however, particles can recognize the target and gather around it (as it can be seen in figure (5) and (6)). Because of clutter we need to apply some preprocessing for clutter suppression (background subtraction), which remove clutter but decreases the SNR.

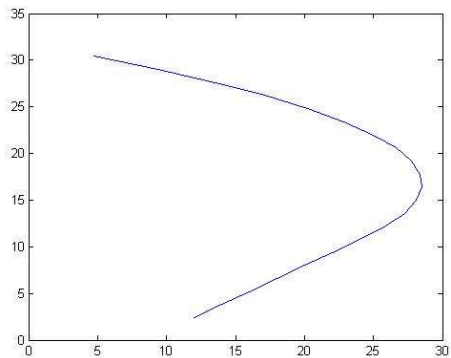

**Fig (3): True trajectory of target**

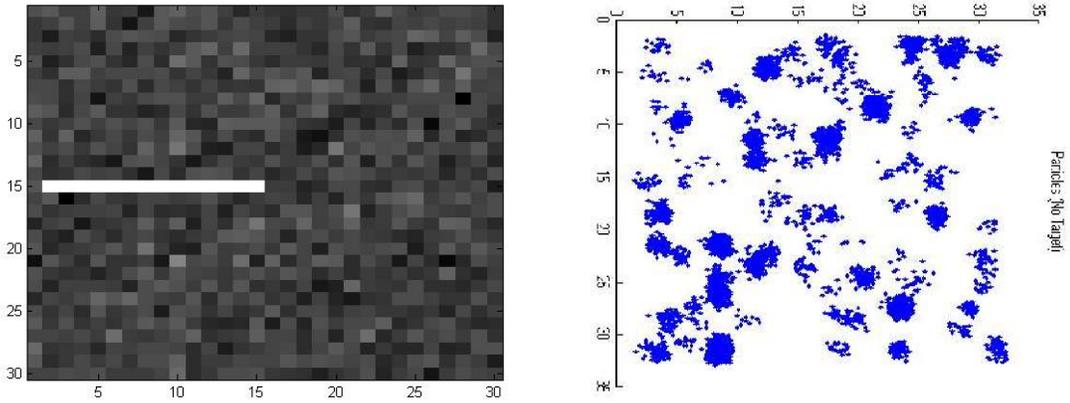

**Fig (4): (left) a frame with clutter without target, (right) particles are uniformly spread**

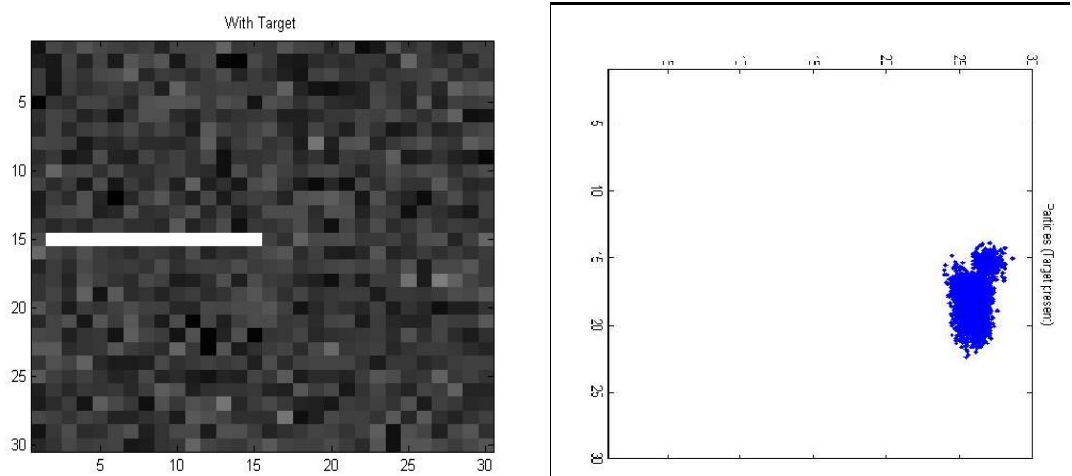

**Fig (5): (left) a frame with clutter with target (it is not visible), (right) particles are centered on target**

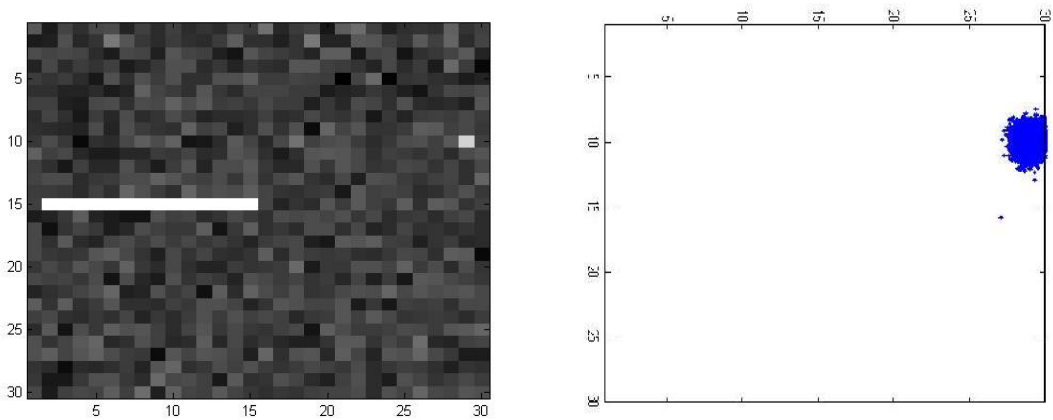

**Fig (6): (left) a frame with clutter with target (it is visible), (right) particles are centered on target**

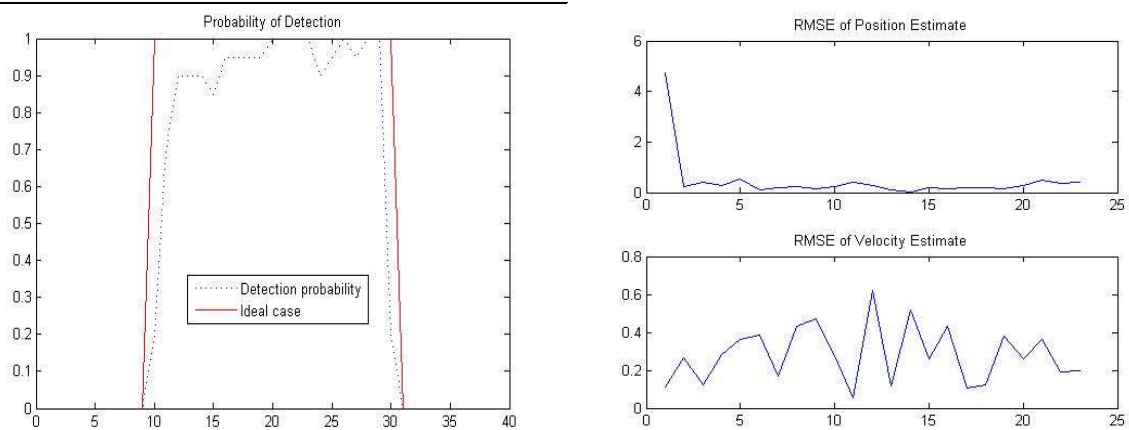

**Fig (7): (left) Detection probability, (right) RMSE for tracking**

## Scenario 2: Detection and tracking of a maneuvering point target in clutter and noise background (real image)

In this scenario the background image is a real one (figure (8)) and the same trajectory as the previous scenario. Mean of target intensity is 7, noise standard deviation is 0.5, and the maximum of clutter intensity is about 4.5 (Figure (9)).

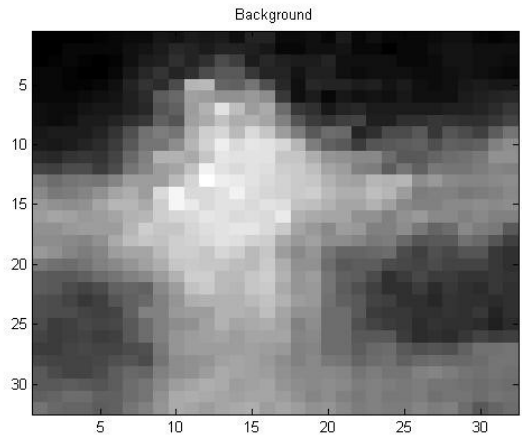

**Fig (8): Background with noise and clutter**

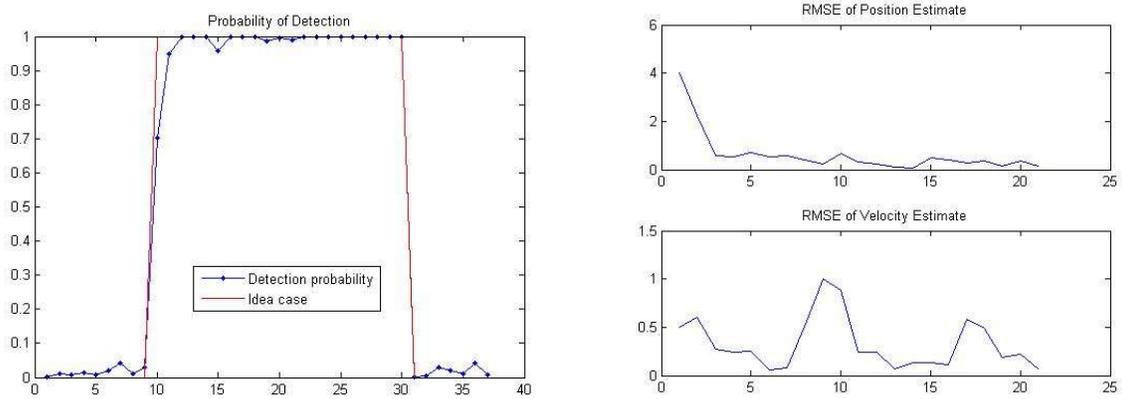

**Fig (9): (left) Detection probability, (right) RMSE for tracking**

The same scenario is considered with a very low SNR (mean of target intensity is 5, standard deviation of noise is 0.5, the maximum intensity of clutter is about 4.5). The results of detection and tracking are as follows (figure (10))

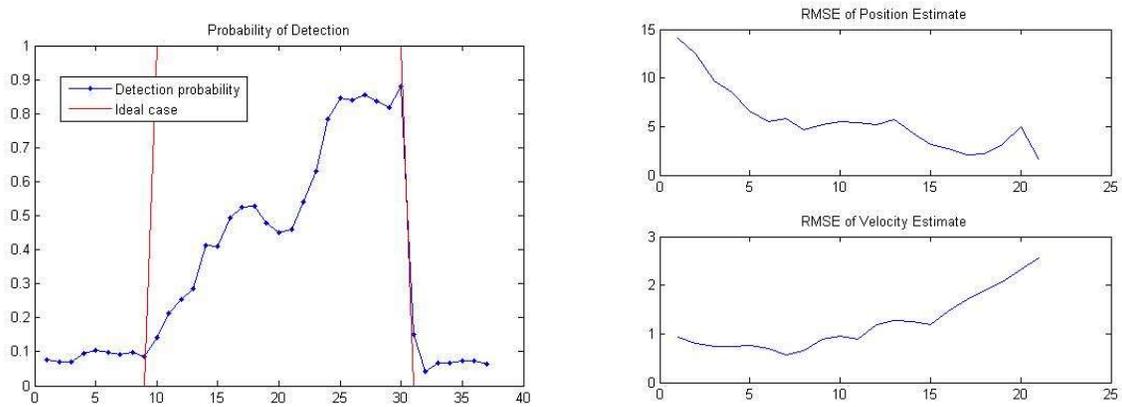

Fig (10): (left) Detection probability, (right) RMSE for tracking

**Scenario 3: Detection and tracking of a maneuvering extended target (without rotation) in clutter and noise background (real image)**

In this scenario the background image is a real one (figure (11)). Mean of target intensity is 20, noise standard deviation is 0.1, and the maximum of clutter intensity is about 1. The results are shown in figure (13).

The same scenario is considered with a very low SNR (mean of target intensity is 10, standard deviation of noise is 0.1, the maximum intensity of clutter is about 1). The results of detection and tracking are as follows (figure (14))

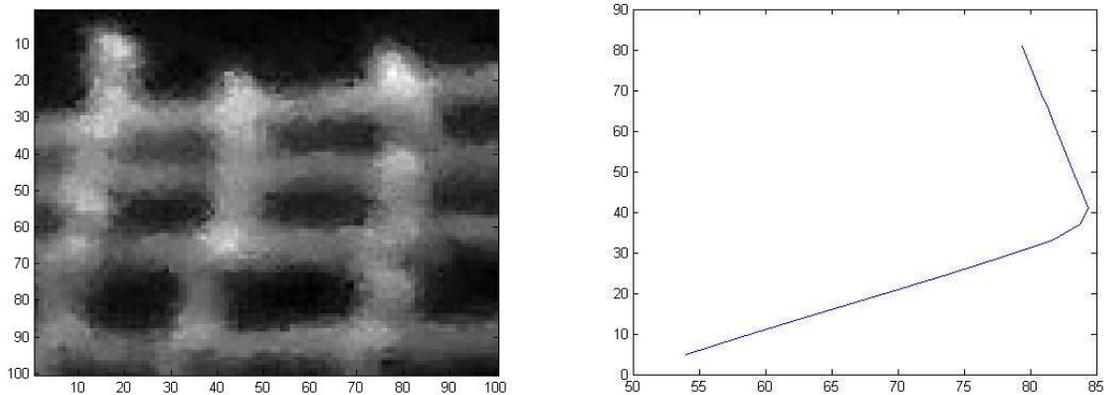

Fig (11): (left) Background with noise and clutter (right) True target trajectory

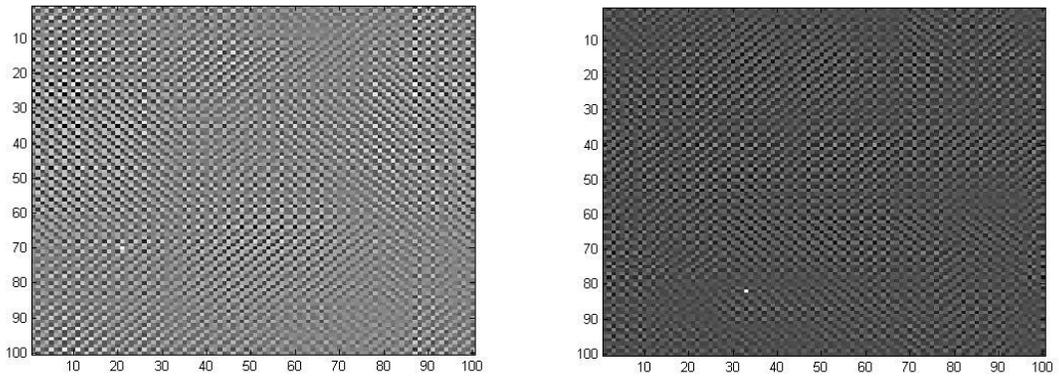

**Fig (12): (left) Frame (including extended target) after preprocessing, (right) Frame (including extended target) after preprocessing**

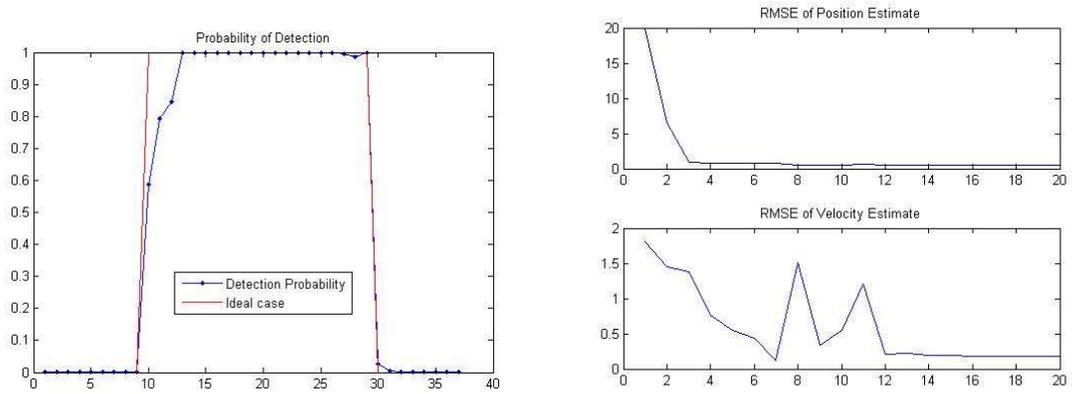

**Fig (13): (left) Detection probability, (right) RMSE for tracking**

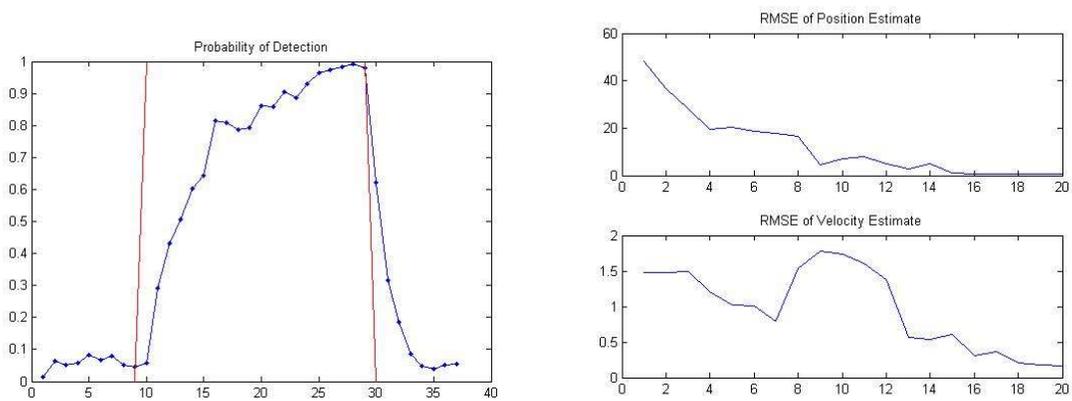

**Fig (14): (left) Detection probability, (right) RMSE for tracking**

**Scenario 4: Detection and tracking of a maneuvering extended target (with rotation) in clutter and noise background (real image)**

In this scenario the background image is a real one (figure (15)) and the trajectory is the same as scenario 1. Mean of target intensity is 6, noise standard deviation is 0.1, and the maximum of clutter intensity is about 1.7. The results are shown in figure (18).

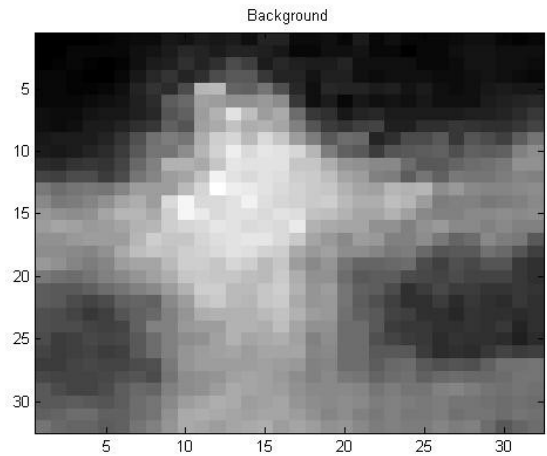

Fig (15): Background with noise and clutter

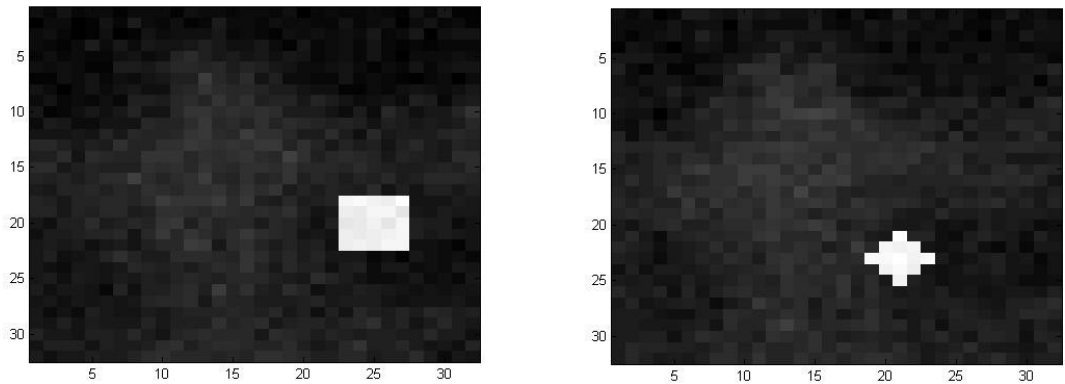

Fig (16): (left) Extended target, (right) Extended target rotated

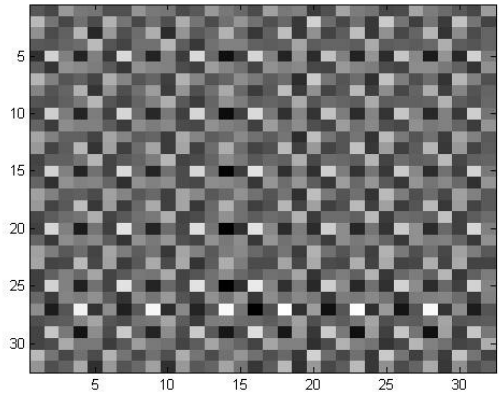

**Fig (17): Frame (including extended target) after preprocessing**

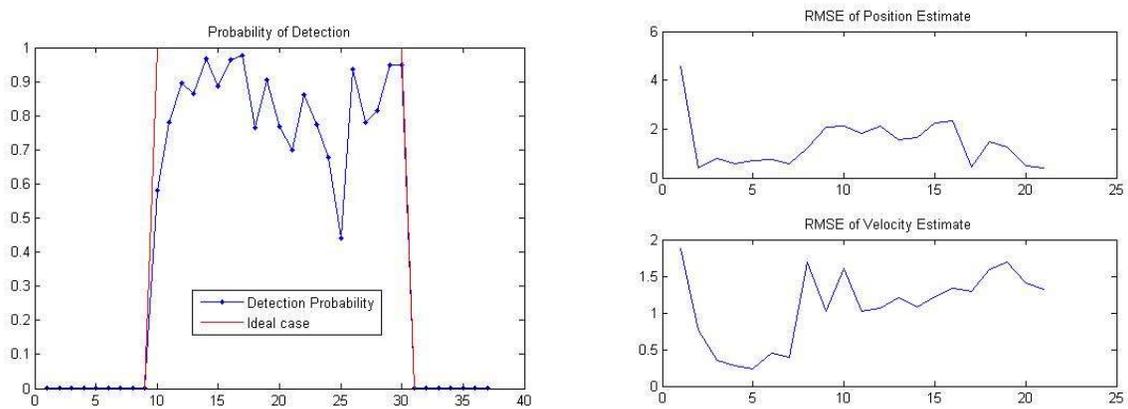

**Fig (18): (left) Detection probability, (right) RMSE for tracking**

## Scenario 4: Detection and tracking of a maneuvering extended target (with rotation) in clutter and noise background with target intensity fluctuations (real image)

In this scenario the background image is a real one (the same as figure (15)) and the trajectory is the same as scenario 1. Mean of target intensity is 9 with uniform fluctuations in the interval of 4 centered at the mean. Noise standard deviation is 0.1, and the maximum of clutter intensity is about 1.7. The results are shown in figure (19).

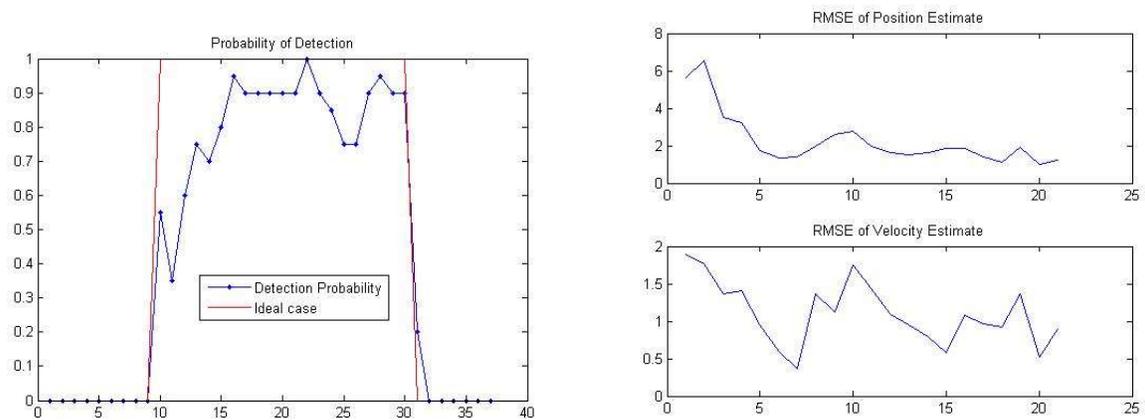

**Fig (19): (left) Detection probability, (right) RMSE for tracking**

## Conclusions and Future Work Directions:

A particle filter-based approach for detection and tracking of a low observable maneuvering point/extended target has been studied. Some preprocessing is required to make the image frames ready for detection and tracking of the target.

Detection and tracking of objects is a decision-estimation problem and, therefore, another approach for handling this problem is based on using a joint decision-estimation approach [16]-[18].

Information about the destination or waypoints for the trajectory of the target has not been incorporated in the above detection and tracking approach. A theoretical foundation of conditionally Markov (CM) sequences was presented in [19]-[26] and their dynamic models, their properties, and some tools were derived for their application to trajectory modeling with destination and waypoint information in [27]-[31]. As a future work, these CM models can be used in the above detection and tracking approach to model and incorporate destination/waypoint information and enhance the detection and tracking performance.

# References

[1]     D. J. Salmond, H. Birch, A particle filter for track-before-detect, *Proceeding of the American Control Conference*, VA, 2001.

[2]     M. Rollason and D. Salmond, "A particle filter for track-before-detect of a target with unknown amplitude," *IEEE Conference,* pp. 14/1-4, 2001.

[3]     M. G. Rutten, N. J. Gordon and S. Maskel, Recursive track-before-detect with target amplitude fluctuations, *IEE Proc. –Radar Sonar Navig.*, Vol 152, No. 5, 2005.

[4]     M. G. Rutten, N. J. Gordon and S. M. Maskel, Recursive track before detect with fluctuating target amplitude, *IEE Proceedings-Radar Sonar navigation,* vol. 152, 2005.

[5]     Branko Ristic, Sanjeev Arulampalam, and Neil Gordon. *Beyond the Kalman Filter: Particle Filters for Tracking Applications*. Artech House, 2004.

[6]     Y. Boers and H. Driessen, "A particle-filter-based detection scheme," *IEEE Signal processing Letters,* vol. 10, pp. 300-302, 2003.

[7]     Y. Boers, H. Driessen, Particle filter based track-before-detect algorithms, *Proceeding of SPIE*, vol. 5204, pp. 20-30, 2003.

[8]     M. G. Rutten, B. Ristic, N. J. Gordon, A comparison of particle filters for recursive track-before-detect, *7th International Conference on information Fusion*, pp. 169-175, 2005.

[9]     S. J. Davey, M. G. Rutten, A comparison of three algorithms for tracking dim targets, *Information, decision and Control,* pp. 342-347, 2007.

[10]    Y. Boers and J. N. Driessen, "Interacting multiple model particle filter," *IEE Proceedings-Radar Sonar Navigation,* vol. 150, pp. 344-349, 2003.

[11]    M. S. Arulampalam, S. Maskell, N. Gordon, and T. Clapp, "A tutorial on particle filters for online nonlinear/non-Gaussian Bayesian tracking," *IEEE Transaction on Signal Processing,* vol. 50, pp. 174-188, 2002.

[12]    Y. Bar-Shalom et al, *Estimation with application to tracking and navigation*, John Wiley, 2001.

[13]    N. J. Gordon, D. J. Salmond, and A. F. M. Smith, "Novel approach to nonlinear/non-Gaussian Bayesian state estimation," *IEE Proceedings-F,* vol. 140, pp. 107-113, 1993.

[14]    A. Doucet, et al, On sequential Monte Carlo sampling methods for Bayesian filtering, *Statistics and Computing*, vol. 10, no. 3, pp. 197-208, 2000.

[15] D. Crisan and A. Doucet, "A survey of convergence results on particle filtering method for practitioners," *IEEE Transaction on Signal Processing,* vol. 50, pp. 736-746, 2002.

[16] X. Rong Li, Optimal bayes joint decision and estimation, *2007 10th International Conference on Information Fusion*, pp. 1-8, 2007.

[17] X. Rong Li, Ming Yang and Jifeng Ru, Joint tracking and classification based on bayes joint decision and estimation, *2007 10th International Conference on Information Fusion*, pp. 1-8. 2007.

[18] R. Rezaie and X. R. Li. Determination, Separation, and Tracking of an Unknown Time-Varying Number of Maneuvering Sources Based on Bayes Joint Decision-Estimation. 18[th] *International Conference on Information Fusion (Fusion)*, Washington, DC, USA, pp. 1848-1855, July 2015.

[19] R. Rezaie and X. R. Li. Gaussian Conditionally Markov Sequences: Dynamic Models and Representations of Reciprocal and Other Classes. *IEEE Transactions on Signal Processing*, vol. 68, pp. 155-169, 2020.

[20] R. Rezaie and X. R. Li. Gaussian Conditionally Markov Sequences: Singular/ Nonsingular. *IEEE Transactions on Automatic Control*, vol. 65, no. 5, pp. 2286-2293, 2020.

[21] R. Rezaie and X. R. Li. Gaussian Conditionally Markov Sequences: Algebraically Equivalent Dynamic Models. *IEEE Transactions on Aerospace and Electronic Systems*, vol. 56, no. 3, pp. 2390-2405, 2020.

[22] R. Rezaie and X. R. Li. Gaussian Conditionally Markov Sequences: Modeling and Characterizations. *Automatica*, vol. 131, 2021.

[23] R. Rezaie and X. R. Li. Nonsingular Gaussian Conditionally Markov Sequences. *2018 IEEE Western New York Image and Signal Processing Workshop (WNYISPW)*, Rochester, NY, USA, pp. 1-5, Oct. 2018.

[24] R. Rezaie and X. R. Li. Explicitly Sample-Equivalent Dynamic Models for Gaussian Conditionally Markov, Reciprocal, and Markov Sequences. *International Conference on Control, Automation, Robotics, and Vision Engineering*, New Orleans, LA, USA, pp.1-6, Nov. 2018.

[25] R. Rezaie and X. R. Li. Models and Representations of Gaussian Reciprocal and Conditionally Markov Sequences. *Proceeding of 2[nd] International Conference on Vision, Image and Signal Processing (ICVISP)*, Las Vegas, NV, USA, pp. 65:1-65:6, Aug. 2018.

[26] R. Rezaie and X. R. Li. Gaussian Reciprocal Sequences from the Viewpoint of Conditionally Markov Sequences. *Proceeding of 2[nd] International Conference on Vision, Image and Signal Processing (ICVISP)*, Las Vegas, NV, USA, pp. 33:1-33:6, Aug. 2018.

[27]     R. Rezaie and X. R. Li. Destination-Directed Trajectory Modeling, Filtering, and Prediction Using Conditionally Markov Sequences. IEEE Transactions on Aerospace and Electronic Systems vol. 57, no. 2, pp. 820-833 ,2021.

[28]     R. Rezaie, X. R. Li, and V. P. Jilkov. Conditionally Markov Modeling and Optimal Estimation for Trajectory with Waypoints and Destination. *IEEE Transactions on Aerospace and Electronic Systems*, vol. 57, no. 4, pp. 2006-2020, 2021.

[29]     R. Rezaie and X. R. Li. Trajectory Modeling and Prediction with Waypoint Information Using a Conditionally Markov Sequence. *56th Allerton Conference on Communication, Control, and Computing (Allerton)*, Monticello, IL, USA, pp. 486-493, Oct. 2018.

[30]     R. Rezaie and X. R. Li. Destination-Directed Trajectory Modeling and Prediction Using Conditionally Markov Sequences. *2018 IEEE Western New York Image and Signal Processing Workshop (WNYISPW)*, Rochester, NY, USA, pp. 1-5, Oct. 2018.

[31]     R. Rezaie and X. R. Li. Mathematical Modeling and Optimal Inference of Guided Markov -Like Trajectory. *2020 IEEE International Conference on Multisensor Fusion and Integration for Intelligent Systems (MFI)*, pp. 26-31, 2020.

# Appendix:

**State Vector Density Function**

Density function can be expanded based on the previous time instance as follows

$$p(x_k|E_k=1,Z^k) = p(x_k|E_k=1,E_{k-1}=1,Z^k)p(E_{k-1}=1|E_k=1,Z^k)$$
$$+ p(x_k|E_k=1,E_{k-1}=0,Z^k)p(E_{k-1}=0|E_k=1,Z^k) \quad (17)$$

The first density in RHS is called survival density and the second one is called new-born density. Then based on Bayes formula

$$p(x_k|E_k=1,E_{k-1}=1,Z^k) = \frac{p(Z_k|x_k,E_k=1)p(x_k|E_k=1,E_{k-1}=1,Z^{k-1})}{p(Z_k|E_k=1,E_{k-1}=1,Z^{k-1})} \quad (18)$$

The numerator and denominator of the equation (18) are divided by $p(Z_k|E_k=0)$ and the likelihood function of the frame based on the likelihood function at each pixel is as follows

$$L(Z_k|x_k,E_k=1) = \prod_i \prod_j l(z_k^{(i,j)}|x_k,E_k=1) \quad (19)$$

Then

$$p(x_k|E_k=1,E_{k-1}=1,Z^k) = \frac{L(Z_k|x_x,E_k=1)p(x_k|E_k=1,E_{k-1}=1,Z^{k-1})}{L(Z_k|E_k=1,E_{k-1}=1,Z^{k-1})} \quad (20)$$

Since a point target just has a contribution to one pixel (or an extend target just has contribution to its neighborhoods), the likelihood function can be simplified for that (those) pixels because other pixels are 1.

The predicted density in the numerator of (20) can be written based on dynamic model as follows

$$p(x_k|E_k=1,E_{k-1}=1,Z^{k-1}) =$$
$$\int p(x_k|x_{k-1},E_k=1,E_{k-1}=1)p(x_{k-1}|E_{k-1}=1,E_k=1,Z^{k-1})dx_{k-1} \quad (21)$$

The second density in (17) can be written as

$$p(x_k|E_k=1,E_{k-1}=0,Z^k) \propto L(Z_k|x_k,E_k=1)p(x_k|E_k=1,E_{k-1}=0) \quad (22)$$

This density is related to a newborn target.

The other term in RHS of (17) can be written based on Bayes formula as follows

$$p(E_{k-1} = 1 | E_k = 1, Z^k) =$$
$$\frac{p(Z_k | E_k = 1, E_{k-1} = 1, Z^{k-1}) p(E_k = 1 | E_{k-1} = 1) p(E_{k-1} = 1 | Z^{k-1})}{p(Z_k, E_k = 1 | Z^{k-1})}$$
$$\propto L(Z_k | E_k = 1, E_{k-1} = 1, Z^{k-1})(1 - p_d) p_{k-1} \quad (23)$$

The first term of RHS is the same as the denominator of (20) which is the normalizing term. So, it can be written as

$$L(Z_k | E_k = 1, E_{k-1} = 1, Z^{k-1}) = \int L(Z_k | x_k, E_k = 1) p(x_k | E_k = 1, E_{k-1} = 1, Z^{k-1}) dx_{k-1} \quad (24)$$

The other term in (17) can be calculated in the same way as follows

$$p(E_{k-1} = 0 | E_k = 1, Z^k) \propto L(Z_k | E_k = 1, E_{k-1} = 0) p_b (1 - p_k) \quad (25)$$

The likelihood in (25) can be written in the same way as follows

$$L(z_k | E_k = 1, E_{k-1} = 0) = \int L(z_k | x_k, E_k = 1) p(x_k | E_k = 1, E_{k-1} = 0) dx_k \quad (26)$$

Therefore, it is possible to recursively estimate the posterior density of target state vector.

**Probability of Target Presence**

In this subsection, probability of target presence is estimated and used as a test statistic. Then, by applying a threshold (e.g., 0.6) on the estimated probability, the algorithm can decide if there is a target in the space or not.

The probability of target presence based on all observations since the beginning to the current time can be estimated as follows

$$p(E_k = 1 | Z_k) = p(E_k = 1, E_{k-1} = 1 | Z_k) + p(E_k = 1, E_{k-1} = 0 | Z_k)$$
$$\propto L(z_k | E_k = 1, E_{k-1} = 1, Z_{k-1})(1 - p_d) p_{k-1} + L(z_k | E_k = 1, E_{k-1} = 0) p_b (1 - p_{k-1}) \quad (27)$$

in which the likelihood terms have already been calculated in the calculation of target state density. For normalizing the probability of (27), the following probability is also calculated

$$p(E_k = 0|Z^k) \propto p_d p_{k-1} + (1 - p_b)(1 - p_{k-1}) \tag{28}$$